\documentclass{aa}
\usepackage{graphicx, natbib}
\bibpunct{(}{)}{;}{a}{}{,}

\begin{document}

\title{An atlas of mid-infrared dust emission in spiral galaxies\thanks{Based
       on observations with ISO, an ESA project with instruments funded by
       ESA Member States (especially the PI countries: France, Germany,
       the Netherlands and the United Kingdom) and with the participation
       of ISAS and NASA.}}

\author{H. Roussel\inst{1}
\and L. Vigroux\inst{1}
\and A. Bosma\inst{2}
\and M. Sauvage\inst{1}
\and C. Bonoli\inst{3}
\and \\
     P. Gallais\inst{1}
\and T. Hawarden\inst{4}
\and J. Lequeux\inst{5}
\and S. Madden\inst{1}
\and P. Mazzei\inst{3}}

\institute{DAPNIA/Service d'Astrophysique, CEA/Saclay, 91191 Gif-sur-Yvette cedex, France
\and Observatoire de Marseille, 2 Place Le Verrier, 13248 Marseille cedex 4, France
\and Osservatorio Astronomico di Padova, 5 Vicolo dell'Osservatorio, 35122 Padova, Italy
\and Joint Astronomy Center, 660 N. A'ohoku Place, Hilo, Hawaii 96720, USA
\and Observatoire de Paris, 61 Avenue de l'Observatoire, 75014 Paris, France}

\offprints{H. Roussel (e-mail: hroussel@cea.fr)}

\date{Received 3 November 2000 / Accepted 23 January 2001}

\maketitle

\begin{abstract}
We present maps of dust emission at 7\,$\mu$m and 15\,$\mu$m/7\,$\mu$m intensity
ratios of selected regions in 43 spiral galaxies observed with ISOCAM.
This atlas is a complement to studies based on these observations, dealing
with star formation in centers of barred galaxies and in spiral disks. It is
accompanied by a detailed description of data reduction and an inventory of generic
morphological properties in groups defined according to bar strength and
H{\scriptsize I} gas content.
\end{abstract}

\section{Introduction}

The mid-infrared maps presented here are part of a larger sample of nearby
spiral galaxies analyzed in Roussel et al. (2001a and b: Papers~I and II).
Paper~I primarily investigates the dynamical effects of bars on circumnuclear
star formation activity, studied through dust emission at 7 and 15\,$\mu$m, and
proposes an interpretation of these mid-infrared data in conjunction with
information on molecular gas and stellar populations. Paper~II deals with
the use of mid-infrared fluxes as a star formation indicator in spiral disks.
Observations belong to guaranteed time programs of ISOCAM, a camera operating
between 5 and 18\,$\mu$m onboard the satellite ISO \citep[described by][]{Cesarsky}.
The sample consists of a first group of large and regular spirals, mainly
barred, and of a second group of spiral galaxies belonging to the Virgo
cluster. The angular resolution of $\approx 10\arcsec$ (HPBW), combined with
spectroscopic information from the same instrument, enables a detailed view of
the variations of two dust phases \citep[unidentified infrared band carriers and
very small grains, as characterized by {\it e.g.}][]{Desert}, tracing
different physical conditions. A stellar contribution can also exist at 7\,$\mu$m
\citep{Boselli}, but is negligible except in a few early-type galaxies.

The general description of the sample and detailed photometric results are
given in Paper~I, where the spectra are also published. Here, we provide the
details of observations and data reduction, total fluxes at 7 and 15\,$\mu$m,
a morphological description of galaxies grouped into appropriate categories
and maps at 7\,$\mu$m, along with optical images. The 15\,$\mu$m maps have very
similar morphology and are therefore not shown. These images demonstrate
the peculiar character of circumnuclear regions seen in the mid-infrared.
Information on the $F_{15}/F_7$ colors of bright complexes and other selected
regions is added.

All of the maps are published for the first time, except those of M\,51
\citep{Sauvage} and NGC\,6946 \citep{Malhotra}, which are both re-analyzed here.

\section{Observations}

All galaxies were observed with two broadband filters, LW3 centered at 15\,$\mu$m
(12--18\,$\mu$m) and LW2 centered at 7\,$\mu$m (5--8.5\,$\mu$m), that we shall
hereafter designate by their central wavelength. Maps covering the whole
infrared-emitting disk were built in raster mode, with sufficient overlap between
adjacent pointings (half the detector field of view in most cases) to avoid border
artifacts and to provide redundancy for spoiled exposures. In all cases,
the field of view is large enough to obtain a reliable determination of
the background level, except for NGC\,4736 and 6744. NGC\,5457, the largest
spiral in the sample with an optical size of nearly $30\arcmin$, was imaged 
with a combination of two overlapping observations made during different
revolutions of the satellite. The pixel size is either $3\arcsec$ or $6\arcsec$,
depending on the galaxy size, but for Virgo galaxies it was always $6\arcsec$, in
order to increase the signal to noise ratio. The half-power/half-maximum diameters
of the point spread function are respectively $6.8\arcsec$/$\simeq 3.1\arcsec$ at
7\,$\mu$m with a $3\arcsec$ pixel size, $9.5\arcsec$/$5.7\arcsec$ at 7\,$\mu$m with
a $6\arcsec$ pixel size, $9.6\arcsec$/$3.5\arcsec$ at 15\,$\mu$m with a $3\arcsec$
pixel size and $14.2\arcsec$/$6.1\arcsec$ at 15\,$\mu$m with a $6\arcsec$ pixel
size. Table~\ref{tab:tab_rasters} summarizes the relevant information regarding
the observations.

To estimate the relative importance of different emitting species in different
galactic regions, spectral imaging between 5 and 16\,$\mu$m of the inner disks
of five bright galaxies was also carried out. These observations are essential
to interpret broadband imaging results. They consist of one pointing with two
circular variable filters, from 16.5 to 9\,$\mu$m and from 9.6 to 5\,$\mu$m,
with a spectral resolution $\lambda/\Delta \lambda {\rm (FWHM)} = 35$ to 50.
The sampling varies between 0.24 and 0.45 of $\Delta \lambda {\rm (FWHM)}$.

\section{Data analysis}

\subsection{Broadband filter maps}

Data reduction was performed using and adapting the ISOCAM
Interactive Analysis package (CIA). The main difficulty and source of
uncertainty is the slow time response of the detector, which is typical of
cold photoconductors. Below 20\,K, the mobility of carriers becomes very slow,
which generates long time constants, both in the bulk of the photoconductor and
at contact electrodes. The typical time spent per pointing
is $\approx 40$--100\,s whereas in general several hundred seconds are needed
to reach stabilization within 10\% of the flux step.

Several causes of memory effects can be distinguished, although they
always involve the same physics of the detector. We will use different
designations for convenience: \\
-- (short-term) transients: they are due to the slow stabilization following a
flux step (either upward or downward), after moving from one sky position to
another; they are the origin of remnant images seen after observing a bright
source. \\
-- long-term transients: drifts and oscillations visible during the
whole observation and influenced by previous history; contrary to short-term
transients, no model exists to correct for them. \\
-- glitch tails: slow return to the sky flux level following the deposition
of energy in the detector by intense cosmic rays (this is an unpredictable
effect). \\
-- responsivity drops: they also follow cosmic ray impacts and produce
``holes'' which can last more than one hundred exposures (this is again an
unpredictable effect).

The data reduction proceeded in the following way: \\
1 -- The subtraction of the dark current follows a model which predicts its time
evolution for even and odd rows of the detector. It accounts for variations along
each satellite orbit and along the instrument lifetime.\footnote{It is described in
``The ISOCAM dark current calibration report'' by Biviano et al. (1998), which
can be found at: http://www.iso.vilspa.esa.es/users/expl\_lib/CAM\_list.html.} \\
2 -- If needed, we mask out the borders of the detector which are not
sufficiently illuminated to allow a proper flat-fielding. \\
3 -- The removal of cosmic ray impacts is complicated by the long-duration memory
effects they can produce. The arrival of an energetic cosmic ray may be followed
by either a slowly decreasing tail or a responsivity drop, as explained above.
The vast majority of glitches are, however, of short duration (one to three
exposures) and can be rejected by median filtering.
To also remove small tails and plateaus induced by glitch pile-up,
the exposures immediately surrounding exposures flagged by the deglitcher were
examined and their flux level compared with that of adjacent non-flagged
exposures, {\it i.e.} we iterated the median filtering once after rejecting the
cosmic ray peaks.
This filtering cannot blindly be applied everywhere. First, at a step between
a faint and a bright illumination level and vice-versa, some real signal variation
may be masked by the deglitcher.
Therefore, the signal in the first and last exposures for each sky
position was compared with that in the neighboring exposures for the same
sky position, and flags cancelled if necessary.
Second, during a pointing on a bright peaked source such as galactic circumnuclear
regions, the noise includes, in addition to the readout and photon noises, high
amplitude fluctuations which are approximately proportional to the signal and
due to the satellite jitter. 
For the purpose of glitch detection, an additional noise component
proportional to the signal was thus tuned to reproduce the jitter effects,
and for very bright sources, defined by a flux threshold set above the
background, fluctuations of 20\% around the median flux were allowed.
The memory effects following some glitches, for which no model exists,
are temporarily set aside and examined at step~6. \\
4 -- Short-term transients are corrected for by using the
latest available technique taking into account the detector characteristics
\citep{Coulais}. The parameters of the model have been determined in low
and uniform illumination cases only, but are likely to vary with signal intensity
and are inexact in the presence of strong spatial gradients. Indeed, when switching
from a moderate flux level to a bright resolved source, the correction hardly
modifies the data.
After a large amplitude flux step, the stabilization is, however, faster,
and thus the last exposures are likely closer to the real flux than in the case
of small flux steps. The readout histories of pixels imaging galactic nuclei
or extranuclear bright complexes were carefully examined after step~7 to check
whether the computed flux level was consistent with the level expected from the
last valid exposures; if not, it was replaced with the mean signal in the last
part of the readout history that was approximately flat as a function of time.
Furthermore, after seeing a bright source, remnant images are not completely
suppressed and can persist for several successive pointings; they are masked
out at step~9. \\
5 -- As long-term transients are not taken into account by the above correction,
they were approximated and removed as one offset for each sky position,
with respect to the last sky position taken as a reference, step by step backwards
and assuming that all pixels follow the same evolution. This is done by comparing
the median of the highest possible number of pixels seeing the off-source sky
in the current pointing and in any of the already-corrected pointings, and selecting
the same set of pixels in both, since at this stage the images have not been
flat-fielded. As it is treated as an offset, the correction does not modify the
source flux but flattens the sky and allows better determination of the background
level. It is similar to the long-term transient correction in the SLICE software
of \citet{Miville}. \\
6 -- Short-term transients after a strong downward flux step which were
imperfectly corrected at step~4, or glitch tails, left aside at step~3, and
with the strongest time derivatives (hence the easiest to detect), were masked
out by an automatic procedure. Drops below the background level, due to a
responsivity drop after a glitch, were also masked out (without the need
for strong time derivatives to design an automatic rejection). \\
7 -- The valid exposures were averaged for each sky position. \\
8 -- The flat-field response was systematically determined for each pixel as its
mean value when illuminated by the background emission (masking the source),
and then normalized by the mean in the $12 \times 12$ central pixels.
When for a given pixel the number of useable values is too low (which
happens if the pixel mostly images the source, or is often rejected due to
glitches or memory effects), or when the error on the mean (computed from
the dispersion) is higher than 5\%, the flat-field response of this pixel is
not computed as the mean, but interpolated using the neighboring pixels and
the default calibration flat-field, before applying the normalization. This
method produces better results than the use of raw calibration files.
These calibration data were employed
only for NGC\,4736 and NGC\,6744, for which the field of view is too small to
apply the above method (the background is hardly seen). For Virgo galaxies, which
are small compared to the camera field of view and were mapped with numerous
pointings, a time-variable flat-field was estimated, improving the final quality
of the maps.
The correction for long-term drifts (step~5) was then iterated once
for all galaxies (this is again similar to the SLICE processing). \\
9 -- The most conspicuous residual memory effects escaping rejection because
they vary too slowly were removed manually in each sky position image.
They were identified visually by comparing structure seen at the same
sky coordinates in different overlapping pointings (corresponding to
different sets of pixels with different histories), taking advantage of the
spatial redundancy of the observations. \\
10 -- The final map was built by projecting together all the individual images
on a sky rectangular grid. \\
11 -- The conversion from electronic units to spectral energy density units is made
using the standard in-flight calibration data base.

\subsection{Spectral imaging}
\label{sec:cvf}

In order to be able to decompose broadband observations according to the
various emitting species, low-resolution spectra between 5 and 16\,$\mu$m were
also obtained toward the inner $3\arcmin \times 3\arcmin$ or
$1.5\arcmin \times 1.5\arcmin$ of five galaxies.
Table~\ref{tab:tab_cvfs} summarizes the parameters of these observations.

For spectral imaging, the principles of data reduction are the same as for
broadband filters. Additional problems encountered in this type of observation
are the following: \\
-- The observation is split into two slightly overlapping parts: from 16.5 to
9.2\,$\mu$m and then from 9.5 to 5\,$\mu$m. The junction between the two circular
variable filters, where the spectral transmission is low, can produce artefacts
shortward of 9.2\,$\mu$m for a few wavelength steps, especially because of
non-stabilization. \\
-- There is a small displacement of the source from one wavelength to another,
the total shift reaching up to two $3\arcsec$-pixels, which is due to the rotation
and change of the circular variable filters. This effect was corrected for by
fitting the shape of the galactic central regions by a gaussian to measure their
displacement on the array (for M\,51, it was not possible and was done manually
at the change between the two segments of the variable filter). \\
-- Bright sources, such as galactic central regions, can create ghost images, due
to reflections inside the instrument; in the case of NGC\,1365, we could identify
and isolate regions contaminated by obvious stray light ghosts. The flux collected
by ghosts in a uniform illumination case is $\approx 20$\% (ISOCAM
handbook)\footnote{ISO handbook volume III: CAM -- the ISO camera, Siebenmorgen
et al., 2000, SAI-99-057. It can be found at:
http://www.iso.vilspa.esa.es/manuals/HANDBOOK/III/cam\_hb/.}.
For resolved sources, it depends on their position on the detector. \\
-- The flat-field response used ignores any time or wavelength dependency.
Nevertheless, this source of error is negligible compared with memory effects and
the error on the determination of the background level. \\
-- Probably due to an inaccurate dark current subtraction
amplified by the transient correction algorithm,
the short-wavelength spectrum can be negative at faint signal levels. \\
-- As sources occupy the whole field of view and no spectra of empty regions were
taken during the observations, the zodiacal foreground produced by interplanetary
dust cannot be directly measured. We estimated it in the following way: we use
a high signal to noise ratio calibration observation, from 16 to 5\,$\mu$m and then
back from 5 to 16\,$\mu$m (which is useful to assess memory effects), of an empty
part of the sky.\footnote{The identification number of this observation, ``TDT
number'' in the ISO archive, is: 06600401.} We apply an offset and a normalization
to this zodiacal spectrum in order to obtain the maximum lower envelope to the
average spectrum of the faintest pixels in our observation. The latter spectrum is
composed of zodiacal light with superimposed faint emission bands (UIBs: see
Paper~I). Then, two bracketing zodiacal spectra with the same normalization are
placed symmetrically; they are constrained to remain inside the dispersion limits
of the observed spectrum, at least where it is not contaminated by UIBs. The
minimum and maximum zodiacal spectra estimated in this way are subtracted from
the galaxy spectrum, which produces an upper and a lower limit to the true
spectrum. The minimum spectrum of the galaxy disk is also required to be positive,
which tightens the bracketing of the zodiacal spectrum.

\begin{figure}[!b]
\vspace*{3cm}
\caption{Spectra of the central regions of the five galaxies observed in
         spectral imaging mode. They have been shifted for clarity by respectively
         0.5, 1.3, 2.1 and 2.9\,mJy.\,arcsec$^{-2}$ and multiplied by the numbers
         indicated on the right. Errors bars include the uncertainty on the
         zodiacal foregroung level, which is small with respect to the central
         source brightness.}
\label{fig:spectres}
\end{figure}

Due to the fact that the zodiacal light is several times higher than the disk
spectrum (outside the bright central regions), the absolute intensity of
the disk emission is subject to large uncertainties. However, since the zodiacal
light spectrum is mainly featureless (it consists of a regularly rising part between
5 and 12\,$\mu$m and a slowly rising plateau from 11 to 16\,$\mu$m: see Fig.~1 in
Paper I), the spectral shape of the disk emission is well determined. \\
-- Because of ghosts, the flat-field response was not handled as in
broadband filter observations. The observed signal is the sum of the source,
the background and their respective ghosts, multiplied by the flat-field response:
$I = (S + b + g(S) + g(b)) \times ff$.
However, the calibration flat-fields are also contaminated by ghosts:
$ff_{\rm cal} = ff \times (1 + g(b)/b)$.
Therefore, we define the corrected signal as:
$I_{\rm corr} = (I - zodi \times ff_{\rm cal}) / ff_{\rm filter}$, where $zodi$
is the zodiacal spectrum estimated as described above and $ff_{\rm filter}$ is a
flat-field response measured with narrow filters, which does not contain any ghost.
We then obtain $I_{\rm corr} = S + g(S)$: only ghosts due to the source remain.

The spectra resulting from the data processing described above are
shown in Paper~I (for central regions, disks, and the zodiacal emission),
where they are discussed. In Fig.~\ref{fig:spectres} we reproduce the spectra
of the galactic central regions, with their noise and uncertainty on the
zodiacal foreground (but no estimation of the errors due to memory effects
was possible). Contribution from stray-light ghosts can be of the order of
10--20\% (see Sect.~\ref{sec:erreurs}).

\subsection{Photometry}
\label{sec:photom}

In some observations, the background is high compared to the source surface
brightness, and it can be affected by inhomogeneities largely due to an incomplete
correction for both short-term and long-term memory effects. The largest possible
number of pixels has to be used, therefore, for a proper determination. In some other
cases, a faint-level emission in spiral arms can be traced out to the map edge
and likely continues outside the field of view. However, this emission is extremely
faint and does not introduce any significant error in integrated fluxes. The only 
exceptions are NGC\,4736 and NGC\,6744, which are larger than the observation
fields of view.
For NGC\,6744, 12\% of the map in the southern part is blank, because the telescope
did not move during 3 of the 16 programmed pointings. The given total fluxes are
therefore lower limits. Other uncertain fluxes are those of the apparent pair
NGC\,4567/68, which has slightly overlapping disks in projection.

For all galaxies, the integrated flux was measured in concentric circles and
traced as a function of the number of pixels, after excluding some sectors of the
map where the galaxy extends to large distances, so as to see as much uncontaminated
background as possible. The background was fitted in the linear part of this curve.

Total fluxes are given in Table~1 and other photometric results (see below)
in Paper~I.

\subsection{Separation of central regions from disks}
\label{sec:sepcnr}

A common feature of dust emission distribution in our galaxies is a very strong
central peak in a region of typically 1 to 3\,kpc. Sub-structures are barely
seen because the angular resolution is of the order of the size of these central
regions, but can sometimes be suspected from variations in the infrared color
$F_{15}/F_7$. We measured fluxes from this central condensation after defining its
radius from the 7\,$\mu$m map. The galaxy center was first determined by fitting a
two-dimensional gaussian on an appropriate zone whenever possible. Otherwise,
it was placed by visual criteria (for three galaxies).
We computed the azimuthally averaged surface brightness profile in elliptical
annuli to compensate for the inclination on the sky, using as far as possible
orientation parameters derived from detailed kinematical analyses, otherwise
from the position angle and axis ratio of outer isophotes. This profile was
decomposed into a central region represented by a gaussian and an inner disk
represented by an exponential. The scale parameters were fitted on parts of the
profile devoid of any obvious structure, such as rings or spiral arms, and each
component was truncated at the radius where the exponential equals the gaussian,
which defined the photometric aperture radius for the circumnuclear region,
$R_{\rm CNR}$.

Dilution effects were corrected for in two ways. The first consists of
approximating central regions by point sources and dividing the fluxes measured
inside $R_{\rm CNR}$ by the integral of the point spread function (PSF) inside the
same aperture. The second one uses an iterative procedure. The image is first
re-sampled on a fine grid (replacing each pixel by smaller pixels all taking the
same value). The brightest pixel of the original sampling is looked for, the PSF
centered on it is subtracted from both images with a small gain to ensure
convergence of the process (5\% here), and is then replaced in the fine-grid image
with a gaussian containing the same flux as the removed PSF, of FWHM about half
the pixel size. This procedure is analogous to the CLEAN algorithm used in radio
astronomy \citep{Hogbom}.
This is repeated until the residual image becomes reasonably uniform.
For central regions not much more extended than the PSF, both estimates
give equal results to a few percent; they differ in cases of a distribution
flatter than the PSF or in the presence of sub-structure, such as a ring.

Fluxes attributable to the pure spiral disk were then defined as the difference
between total fluxes and central region fluxes measured in this way.

\subsection{Errors and photometric consistency}
\label{sec:erreurs}

Photometric errors cannot be computed rigorously, due to the uncontrolled
memory effects of the camera. The quoted uncertainties are therefore only
indicative. They are the sum of three components:
\begin{list}{--}{}
   \item the readout and photon noises added quadratically, computed from the
   readout history of each pixel;
   \item the error on the background level fitted to the histogram of off-source
   pixels (which includes uncertainties due to residuals of dark current,
   flat-field, glitches, remnant images and long-term drifts), also added
   quadratically;
   \item the deviation from stabilization, grossly estimated from the dynamics
   of each pixel's response during the time spent on each pointing, as half the
   difference between the smoothed variation amplitude and the 99\% confidence
   level readout and photon noise. This error component is added linearly
   over all the pixels imaging the galaxy, because memory effects cannot be
   assimilated to white noise.
\end{list}
The latter error is usually dominant, except for the faintest galaxies.
We obtain typical errors of $\approx 10$\% at 7\,$\mu$m and 18\% at 15\,$\mu$m
(most galaxies having errors less than, respectively, 18 and 30\%).
These do not take into account flux density calibration uncertainties,
which are of the order of 5\%, with an additional 5\% variation along each
orbit (ISOCAM handbook).

The five galaxies observed in spectral imaging mode provide the opportunity to
perform checks on the agreement between fluxes derived from independent
observations. Since the wavelength coverage of our spectra (5 to about 16\,$\mu$m)
does not contain the whole 15\,$\mu$m band (12--18\,$\mu$m), our comparison is
limited to the 7\,$\mu$m band. We simulated observations in the LW2 band
(5--8.5\,$\mu$m) using its quantum efficiency curve. The comparison with the
photometry performed on the 7\,$\mu$m maps is shown in Fig.~\ref{fig:comp_cvf_lw2}.
The agreement is perfect for NGC\,1365, but the spectra of the other four galaxies
systematically overestimate 7\,$\mu$m fluxes with respect to raster maps. The
amplitude of the deviation (23\% for NGC\,613, 16\% for NGC\,1097, 12\% for
NGC\,5194 and 17\% for NGC\,5236) is of the order of what can be expected from
stray light ghosts from the circumnuclear regions, which would be superimposed
onto the source. This is consistent with the fact that an obvious ghost can be
identified (outside the central regions) only in the case of NGC\,1365, whose
circumnuclear region is thus likely free from the ghost it generates. It should
then be remembered that photometry performed on raster maps is more reliable. See
also \citet{Forster} for a photometric comparison between ISOCAM and ISOSWS spectra.

\begin{figure}[!t]
\vspace*{3cm}
\caption{7\,$\mu$m fluxes simulated from the spectra (ordinate) versus 7\,$\mu$m
         fluxes measured in maps (abscissa). The apertures used are centered on
         the galactic nuclei and encompass the entire circumnuclear regions
         (chosen diameters of respectively $24\arcsec$, $48\arcsec$, $48\arcsec$,
         $84\arcsec$ and $48\arcsec$ for NGC\,613, 1097, 1365, 5194 and 5236).
         Note that the nucleus of NGC\,5236 is slightly saturated in the 7\,$\mu$m
         map. Error bars include deviation from stabilization for raster maps
         and the uncertainty on the zodiacal foreground level for spectra.
         The solid line indicates $y = x$ and the dashed line $y = 1.2\, x$.}
\label{fig:comp_cvf_lw2}
\end{figure}

\begin{figure}[!t]
\vspace*{6cm}
\caption{Comparison of our photometric results at 15\,$\mu$m (top) and 7\,$\mu$m
         (bottom) with those of \citet{Dale} (DSH). Our error bars include the three
         components described above but not the systematic flux density calibration
         uncertainty, which acts equally on both sets of data. The solid line
         indicates $y = x$ and the dotted lines $y = (1 + \alpha)\, x$, with
         $\alpha$ varying by steps of 0.1\,.}
\label{fig:comp_dale}
\end{figure}

Second, we can compare our results with independent measurements from the same
ISOCAM raster observations (differing in the applied data reduction and photometry),
for some galaxies of the sample analyzed by other individuals. In the {\it Cambarre}
and {\it Camspir} samples, there is only one such galaxy, M\,83, bright and very
extended, for which \citet{Vogler} give $F_{15} = 20.2$\,Jy and $F_7 = 19.3$\,Jy,
that is to say respectively 1.005 and 1.04 times the values obtained by us: the
difference is well within the error bars. The fluxes of Virgo galaxies from
\citet{Boselli} have not yet been published, so that we cannot check their
consistency with ours.

We have also processed the 7 and 15\,$\mu$m maps of several galaxies from the
{\it Sf\_glx} observation program (PI G. Helou), selected from the ISOCAM
public archive and added to our sample for the analysis presented in
Paper~I. The maps are not presented here since they have been published
in \citet{Dale} (DSH). The results of our own photometry are given in Paper~I.
Fig.~\ref{fig:comp_dale} shows the confrontation of our fluxes with those of
DSH. NGC\,6946, which is brighter by one order of magnitude than the other
galaxies, is not included, for clarity (DSH obtain fluxes within respectively
8\% and 4\% of ours at 15 and 7\,$\mu$m for this object). As seen in
Fig.~\ref{fig:comp_dale}, whereas the agreement at 15\,$\mu$m is reasonable
(for all galaxies except three, the difference is less than 15\%), the results
at 7\,$\mu$m are dramatically discrepant, our fluxes being systematically higher
than those of DSH. As a result, $F_{15}/F_7$ colors computed by DSH are always
higher than ours. To investigate the origin of this discrepancy, we
performed some tests on NGC\,986, an extended galaxy with bright circumnuclear
regions, for which $F_{15}~{\rm (DSH)} = 1.02\, F_{15}~{\rm (us)}$ and
$F_7~{\rm (DSH)} = 0.64\, F_7~{\rm (us)}$.

We substituted each step of our data reduction successively, leaving all the
other steps unchanged, according to what we thought DSH had done from the
information they give on their data processing: use of calibration files
for the dark current; use of the deglitcher described by \citet{Starck} with
default parameters and without any subsequent masking; use of the ``fit3''
routine with default parameters for short-term transient correction; suppression
of the long-term transient correction; and use of calibration files for the
flat-field.

We would advise against the empirical ``fit3'' routine for photometric
purposes (it always finds a solution and artificially produces corrected readout
histories which are flat, but this does not ensure that it converges toward the
right value). Thus we initially thought that photometric discrepancies would be
attributable to the short-term transient correction. However, none of the changes
listed above is sufficient to account for the large difference in the 7\,$\mu$m
fluxes. We also tried to reproduce the photometric method of DSH, but this
also can be excluded as the major source of the discrepancy.

As a last possibility, DSH mention that besides using calibration data, they
also built flat-fields using the observations themselves, by computing 
the median of the images -- but apparently without previously masking the
areas where the galaxy falls --; we do not know to which galaxies exactly this
applies and in which instances they have instead used calibration files. We
have thus explored this track and created such a flat-field for NGC\,986. We find
that this is the only way we can recover the 7\,$\mu$m flux given by DSH.

However, we emphasize that flat-fields should not be built from the observations
without masking the source, especially for an extended galaxy with bright central
regions such as NGC\,986, on a small raster: when doing so, one strongly
overestimates the flat-field response at all the places where bright sources
are observed (for the $2 \times 2$ raster of NGC\,986 for instance, the nucleus
is seen at four places on the detector, and still at three places if the first
sky position is removed, as DSH have done). As a consequence, after dividing by
such a ``flat-field'', the final flux of the galaxy is severely underestimated.
If indeed DSH have proceeded this way, this could explain why their fluxes are
systematically lower than ours. If the same processing holds for 15\,$\mu$m maps,
we speculate that the discrepancy with our 15\,$\mu$m fluxes is reduced due to a
compensation effect by the background: at 15\,$\mu$m, the zodiacal emission is much
higher than at 7\,$\mu$m with respect to the galaxy flux (by a factor 5--6). Since
the contrast between the background and the galaxy is much lower, the quality
of the flat-field response is less affected than at 7\,$\mu$m.

As a further check, one of us (M. Sauvage) independently reduced the maps
and measured the fluxes of NGC\,986 by a method different to that explained in
Sect.~\ref{sec:photom}. At 7\,$\mu$m, allowing the isophotal detection limit to vary
between 0 and 3 times the background dispersion above the background level, the
resulting flux is within $- 6$\% and $+ 8$\% of our tabulated flux, and the
preferred measurement in view of the extension of the galaxy is at $- 4$\%.
At 15\,$\mu$m, these numbers are respectively $- 14$\% and $+ 13$\%, and $- 7$\%.
Hence, our various estimates are in good agreement with each other.

\section{Maps}

For each galaxy are shown, from top to bottom: an optical map
from the Digitized Sky Survey (with a better contrast inset from the second
generation survey, if available and if the large-scale image is saturated);
the 7\,$\mu$m map, to which we applied a transfer function $I^n$, where
$0.4 \leq n \leq 1$, so as to make both faint and bright
structures visible with the grey scale; $F_{15}/F_7$ colors of a few regions,
selected for being regular and bright at 7 or 15\,$\mu$m, which are shown as an
illustration of the constancy of colors in disks, except in a few resolved star
formation complexes and in circumnuclear regions. For these color measurements,
7\,$\mu$m maps were first convolved to the 15\,$\mu$m angular resolution.
As the astrometry is not more precise than $\approx 10\arcsec$ (the absolute
accuracy of ISO, 1--$2\arcsec$, is degraded by ISOCAM's lens wheel jitter),
the displacement between both maps was measured on central regions, by fitting
them with a gaussian when possible. For some Virgo galaxies not observed at 7 and
15\,$\mu$m in concatenated mode but during different revolutions, there is a
rotation between the two maps of up to $5\degr$, which was estimated by fitting
the few usable regular complexes with gaussians.
Apertures were fixed by the angular resolution:
we chose $12\arcsec$ for observations with a $3\arcsec$ pixel size, and $18\arcsec$
for a $6\arcsec$ pixel size, {\it i.e.} $\approx 1.25$ times the half-power
diameter of the point spread function.
We first give the colors of the circumnuclear region inside the radius defined as
explained in Sect.~\ref{sec:sepcnr} (marked `C'), and of the whole averaged disk
(marked `D'). Regions in the bar are noted `B' and regions in arms, arcs or rings
are noted `A'. In some cases, the color of the central region inside the resolution
unit is also given (`Cr'). Fluxes are expressed in mJy, so that to obtain a true
ratio of powers emitted inside the 12--18\,$\mu$m and 5--8.5\,$\mu$m bandpasses
(and not a flux density ratio), one should multiply $F_{15}/F_7$ by $\approx 0.42$.

Errors on average disk colors include only photon and readout noises and the
uncertainty on the background level; for all other color measurements, they
additionally contain an estimate of the errors due to incomplete stabilization.
We again warn the reader that the latter error component cannot be derived
rigorously: what follows is an attempt to provide an order-of-magnitude
estimate.
Since transients are of a systematic nature (the response of the camera to a given
input is perfectly reproducible) and tend to behave coherently in a well defined
region (a brightness peak or minimum), we treated them in the following way: \\
   -- We call $m_i^{\lambda}$ the estimated uncertainty due to memory
   effects in the pixel $i$ of the final map at the wavelength $\lambda$,
   derived as explained in the begining of Sect.~\ref{sec:erreurs}.
   $R^{\lambda}$ is the ensemble of pixels belonging to the region of interest.
   Then the stabilization error on the true total flux $F^{\lambda}$ of that
   region is
   \begin{eqnarray}
      M^{\lambda} = \sum_{i \in R^{\lambda}} m_i^{\lambda}.
   \nonumber
   \end{eqnarray}
   -- We consider two cases: either both resulting errors at 7 and 15\,$\mu$m
   are treated positively, {\it i.e.} fluxes are overestimated (we measure
   $F^{\lambda}_{mes} \approx F^{\lambda} + M^{\lambda}$ instead of $F^{\lambda}$),
   or both are treated negatively, {\it i.e.} fluxes are underestimated
   ($F^{\lambda}_{mes} \approx F^{\lambda} - M^{\lambda}$). We assume the same
   trend in both bands because memory effects are systematic in nature, and we
   consider it equally probable that fluxes are overestimated or underestimated in
   the case of an upward flux step as well as in the case of a downward step,
   because the detector response can oscillate as a function of time. \\
   -- An asymmetric error bar on the color is then directly derived. Let us call
   ${\cal M}^c_+$ and ${\cal M}^c_-$ the bounds of this error bar: the measured
   color is noted
   $\left( F^{15}_{mes} / F^7_{mes} \right)^{+ {\cal M}^c_+}_{- {\cal M}^c_-}$.
   If true fluxes are overestimated,
   \begin{eqnarray}
      \frac{F^{15}_{mes}}{F^7_{mes}} &=& \frac{F^{15}}{F^7} \times
      \frac{1 + M^{15} / F^{15}}{1 + M^7 / F^7}
   \nonumber \\
      &\approx& \frac{F^{15}}{F^7} \times
      \left( 1 + M^{15} / F^{15} - M^7 / F^7 \right)
   \nonumber
   \end{eqnarray}
   using a limited development.
   If $D^c_p$ is the deviation from the true color
   ($F^{15}_{mes} / F^{7}_{mes} - F^{15} / F^7$) induced by an
   overestimation of fluxes and $D^c_n$ the deviation resulting from an
   underestimation of fluxes, one obtains
   \begin{eqnarray}
      D^c_p = \frac{F^{15}_{mes} - M^{15}}{F^7_{mes} - M^7} \times \left[
      \frac{M^{15}}{F^{15}_{mes} - M^{15}} - \frac{M^7}{F^7_{mes} - M^7} \right]
   \nonumber
   \end{eqnarray}
   and similarly
   \begin{eqnarray}
      D^c_n = \frac{F^{15}_{mes} + M^{15}}{F^7_{mes} + M^7} \times \left[
      ~- \frac{M^{15}}{F^{15}_{mes} + M^{15}} + \frac{M^7}{F^7_{mes} + M^7} \right].
   \nonumber
   \end{eqnarray}
   -- There are four possibilities: \\
   If $D^c_p > 0$ and $D^c_n < 0$ then ${\cal M}^c_- = D^c_p$ and
   ${\cal M}^c_+ = - D^c_n$. \\
   If $D^c_p < 0$ and $D^c_n > 0$ then ${\cal M}^c_+ = - D^c_p$ and
   ${\cal M}^c_- = D^c_n$. \\
   If $D^c_p > 0$ and $D^c_n > 0$ then ${\cal M}^c_- = max[D^c_p, D^c_n]$
   and ${\cal M}^c_+ = 0$. \\
   If $D^c_p < 0$ and $D^c_n < 0$ then ${\cal M}^c_+ = - min[D^c_p, D^c_n]$
   and ${\cal M}^c_- = 0$.

~ \\
Maps are organized as follows: \\
Fig.~5: strongly barred galaxies arranged in order of decreasing
        $D_{\rm bar}/D_{25}$, the ratio
        of the deprojected bar diameter to the disk diameter; \\
Fig.~6: galaxies with weak or no bar arranged in the same order; \\
Fig.~7: the peculiar galaxies; \\
Fig.~8: all galaxies of the Virgo cluster sample whatever the bar strength
        which are not HI-deficient; \\
Fig.~9: HI-deficient Virgo galaxies. \\
Maps are oriented with north to the top and east to the left.

\section{Morphological properties}
\label{sec:morpho}

In the following, we grouped galaxies according to their most striking
property. The well-resolved regular spirals of the first subsample were arranged
in order of decreasing $D_{\rm bar}/D_{25}$, the ratio of the deprojected bar
diameter to the disk major diameter at the blue isophote 25 mag.arcsec$^{-2}$.
Virgo galaxies are resolved with less detail, are more frequently highly inclined
and intrinsically smaller and fainter, not allowing a meaningful measure of
$D_{\rm bar}/D_{25}$. They are separated into two groups differentiated by
interactions with the cluster environment: galaxies with a normal H{\scriptsize I}
content and H{\scriptsize I}-deficient galaxies.

Beyond the complexity and distinctive features of all galaxies shown here,
we choose to emphasize in this section the generic properties found within each
group and do not describe in depth each individual object.
The morphological types from the RC3 \citep{Vaucouleurs} are indicated.

We determined the size of the infrared emitting disk (at 7\,$\mu$m because
the sensitivity is better than at 15\,$\mu$m), defined as the diameter of the
isophote $5\,\mu$Jy.\,arcsec$^{-2}$. This is the deepest level that can be reasonably
reached for all galaxies of the sample, corresponding to between one and five times
the background $1 \sigma$ uncertainty (except for NGC\,5194 which has a higher
background noise). This level is very similar to the sensitivity reached by
\citet{Rice} in 12\,$\mu$m IRAS maps. The measure was performed on azimutally
averaged surface brightness profiles, taking into account the orientation and
inclination of the galaxy. We cannot see any difference in the mid-infrared size
to optical size ratio between barred and non-barred galaxies, but this ratio tends
to increase from early-type to late-type galaxies. More fundamental systematics
exist as a function of gas content, which will be described in Sect.~\ref{sec:sizes}.
Indeed, the likely reason for the observed dependence on Hubble type is the
existence of a classification bias for H{\scriptsize I}-deficient galaxies: they
are mostly classified as early type, due to the abnormal aspect of their spiral
arms, and since the bulge-to-disk ratio criterion plays a minor role in the
classification compared to the resolution of spiral arms into star formation
complexes.

\subsection{Well resolved galaxies}

\begin{list}{--}{\itemsep 2ex
                 \leftmargin 0em
                 \itemindent 2ex
                 \parsep 1ex
                 \listparindent 3ex}

\item {\bf Strongly barred spirals:}

NGC\,5383 (SBb), 7552 (SBab), 1433 (SBab), 1530 (SBb), 613 (SBbc), 1672 (SBb),
1097 (SBb) and 1365 (SBb).

The circumnuclear region is extremely bright in the mid-infrared -- always
brighter than any complex in the bar or the arms. This is particularly
obvious in NGC\,7552, whose bar is relatively faint and where only a hint
of the eastern arm is present, whereas emission from the large central region
is very intense. All galaxies of this category are known to possess a nuclear spiral
or ring, and sometimes additionally a nuclear bar. All except NGC\,1433 have a
hot-spot nucleus, {\it i.e.} giant star formation complexes within the central
structure. We give here a brief summary of nuclear morphologies: \\
-- NGC\,5383: nuclear spiral \citep{Buta2} with hot spots \citep{Sersic2}. \\
-- NGC\,7552: nuclear ring \citep{Feinstein} with hot spots -- although the nucleus
   is classified amorphous by \citet{Sersic1}, bright complexes are clearly visible
   in the radio continuum maps of \citet{Forbes} --; \citet{Dottori} infer from a
   population synthesis of stellar absorption lines a cycle of three or four
   successive starbursts. \\
-- NGC\,1433: nuclear ring and nuclear bar \citep{Buta1}, amorphous nucleus
   \citep{Sersic1}. \\
-- NGC\,1530: nuclear spiral \citep{Buta2} with hot spots \citep{Sersic2}. \\
-- NGC\,613: nuclear spiral and nuclear bar \citep{Jungwiert}, hot spots
   \citep{Sersic2}. \\
-- NGC\,1672: nuclear ring with hot spots \citep{Storchi}, although \citet{Sersic1}
   include it among amorphous nuclei. \\
-- NGC\,1097: circumnuclear starburst ring of diameter $20\arcsec \approx 1.5$\,kpc
   with hot spots, nuclear bar \citep{Friedli}. The ring is well resolved in the
   mid-infrared. \\
-- NGC\,1365: nuclear spiral and nuclear bar \citep{Jungwiert}, hot spots
   \citep{Morgan}.

The bar is detected in all galaxies but NGC\,1433, which, besides its
central region, displays only very faint and patchy emission; in NGC\,1365,
it is detected but extremely diffuse.

We also note that the mid-infrared emission of the bar generally consists of
thin bands (as opposed to the stellar bar which is more
oval), shifted towards the leading edge (if one assumes that spiral arms are
trailing), in agreement with the response of other gas tracers to the bar
potential. However, these bands do not coincide strictly with dust lanes seen
in optical images: dust which is conspicuous in optical starlight absorption
is not the same as dust seen in mid-infrared emission. In fact, absorbing dust
lanes are signatures of gas density enhancements, at the location of shocks
\citep{Athanassoulab}.
As these shocks are strongest in the inner parts of the bar, near the central
region, absorption dust lanes are also better developed and show higher contrast
in the inner bar, independently of the brightness of the underlying stellar
continuum (this is particularly well seen in NGC\,1530 and 1672); conversely,
the infrared emission is systematically brightest in the outer parts of the
bar and diffuse near the circumnuclear region. The same argument excludes
significant collisional excitation of dust in shocks: grains are heated by
photons, including in regions of large-scale shocks. \\
A further sign of the different nature of absorption and emission
dust lanes is the aspect of the bar of NGC\,1365: whereas absorption dust
lanes show high contrast and are displaced far from the bar major axis toward
the leading edge, the infrared emission is extremely diffuse and rather
symmetric about the major axis.

Star formation complexes are generally seen in the bar at varying places,
between mid-distance from the nucleus to the end of the bar (see for instance
the bright knots in NGC\,7552 and 1530). Such complexes inside the bar are also
commonly found in H$\alpha$ observations \citep{Garciac, Martina}. Emission is
always enhanced at the junction between the bar and spiral arms, even in NGC\,1433.
This has been interpreted as an effect of orbit crowding, since it is the
place where elongated orbits subtended by the bar meet near-circular orbits
of the disk \citep{Kenney}. This region therefore favours star formation,
unlike the inner bar where the high-velocity shocks are likely to prevent it
\citep{Tubbs, Athanassoulab, Reynaud}.

In some galaxies (NGC\,1530, 1365 and to a lesser extent NGC\,5383), spiral arms
are clearly seen to begin in advance of the bar, {\it i.e.} to continue beyond
the junction with the bar towards the leading edge, which is never the case
for weakly barred galaxies. This could be due to a decoupling between
the bar and spiral waves, {\it i.e.} different pattern speeds. Spiral arms
generally make sharp angles with the bar (except the eastern arms of
NGC\,7552, 613 and 1672), which is also characteristic of strong bars, as
already noted by \citet{Prendergast}.

The inner parts of arms show in most cases much star formation, as evidenced
by bright emission and knots both in the optical and in the mid-infrared
(exceptions are NGC\,7552, whose western arm is not detected at all and
whose eastern arm is hardly seen, and NGC\,1433, whose entire inner ring is
seen near the detection limit. Nevertheless, there seems to be a maximum
radius encompassing sites of reasonable star formation rates, outside of
which spiral arms abruptly become fainter, both in optical and infrared
images. The clearest and most symmetric example is NGC\,1365, but this remark
is valid for all. The exhaustion of outer arms could be a signature of
the corotation between the spiral wave and the gas, but it can also
be related to the fact that all the strongly barred galaxies presented here
are of early type (between SBab and SBb, with one SBbc, NGC\,613). This is not
a bias in the sample: long bars are found predominantly among early types
\citep{Athanassoulaa, Martinb}.

We finally note the presence of an inner plateau or lens of diffuse infrared
emission delineating a sort of inner pseudo-ring in all galaxies of this
group, again except NGC\,7552 and 1433. In NGC\,1530, the plateau is also
seen in diffuse H$\alpha$ emission and the pseudo-ring materialized by H$\alpha$
knots \citep{Regan}. In NGC\,613, 1672 and 1097, a real lens is seen in the
optical and further outlined by secondary tightly wound arms or incomplete
arcs, best seen in the infrared.

\item {\bf Weakly or non barred spirals:}

NGC\,5236 (SABc), 6946 (SABcd), 4535 (SABc), 6744 (SABbc), 289 (SBbc),
5457 (SABcd), 4736 (SAab), 5194 (SAbc).

In weakly barred spirals, the central region is bright in the mid-infrared,
but contrary to strong bar centers, it is relatively small and can be rivaled
in brightness by complexes in the arms. The centers of NGC\,5236 and 6946 are
particularly intense. Several of these galaxies also contain nuclear rings,
but none seems to harbor hot spots comparable to those seen in galaxies of the
first group. Here is a list of nuclear morphologies: \\
-- NGC\,5236: nuclear ring \citep{Buta2}, amorphous nucleus \citep{Sersic1}. \\
-- NGC\,6946: nuclear bar \citep{Zaritsky}. \\
-- NGC\,4535: nuclear ring \citep{Buta2}. \\
-- NGC\,6744, 289 and 5457: no remarkable structure. \\
-- NGC\,4736: weak nuclear bar \citep{Mollenhoff}. \\
-- NGC\,5194: nuclear ring \citep{Buta2}, which is seen in the mid-infrared.

The bar is detected in all SABs except NGC\,6744, which is completely devoid
of infrared emission between the nucleus and the inner ring, much like NGC\,1433.
Unlike strong bars, weak bars do not contain infrared knots: their emission is
rather unstructured. They consist of thin lanes displaced towards the leading edge,
just as in strong bars, but are more curved and smoothly connected to spiral arms,
in agreement with the notations of \citet{Prendergast}.

In NGC\,6946, the bar is rather a fat oval distribution of stars, inside
which a spiral arm can be traced at the north.
The bar that we see in the infrared is shorter and rotated by as much as
$\approx 35\degr$ to the leading side of the stellar oval. It does not
coincide with the small northern arm but corresponds well
to the $\approx 1\arcmin$ molecular bar-like distribution \citep{Ball, Regan2},
confirmed kinematically by \citet{Bonnarel}, and in which
is embedded a nuclear bar rotated by a further angle of $\approx 15\degr$
to the leading side.

Zones at the junction between the bar and spiral arms are much less conspicuous
than in strongly barred galaxies (except the SE junction in NGC\,289 which is bright).
Brightness peaks are found further out along the arms (see in particular
NGC\,4535). Dust emission follows remarkably closely the spiral structure at very
large distances from the center. Note for instance the strong similarity between
optical and infrared images of NGC\,4535, where even the slight brightness enhancement
at the northern edge of the disk, likely due to compression by the intracluster gas,
is detected at 7 and 15\,$\mu$m. There is no sharp brightness decrease beyond some
radius as in strongly barred spirals, except in NGC\,289. This galaxy appears to be
the closest of its category to strongly barred galaxies (bright knot at the end of
the bar, rather sharp angle of spiral arms with the bar, slight advance of the arms
on the bar, abrupt decrease in surface brightness between an inner disk and an outer
disk), despite its rather short bar. However, we note that the outer disk, whose
size defines the normalization to the bar size, is extremely tenuous, unlike in
the other spirals of this group.

NGC\,4736, although classified SA, has an oval distorsion revealed by its
velocity field \citep{Bosma}, hence a weak bar. Its metallicity radial
gradient is typical of SAB galaxies and could perhaps result from the past
mixing effects of a stronger bar now dissolved \citep{Martinc}.

\item {\bf Peculiar spirals:}

NGC\,1022 (SBa) and 4691 (SB0/a).

NGC\,4691 is described as an ``irregular galaxy of the M\,82 type''
by \citet{Sandagea} and then as amorphous \citep{Sandageb}, based on the
chaotic, ``intricate and extensive dust pattern in the central regions'' which
is similar to that in some starburst galaxies such as M\,82 and suggests
``galactic fountain activity''. This interpretation is supported by the
spectroscopic data of \citet{Garciaa}, whose I-band and H$\alpha$ images also
reveal a peculiar central structure with four knots. The two strongest H$\alpha$
knots are resolved in our mid-infrared images, and are surrounded by two
concentrations of molecular gas aligned with them, but at nearly twice the
distance from the center \citep{Wiklind}. In the mid-infrared, we detect
only the central structures and not the bar or the rest of the disk.

NGC\,1022 is also peculiar and resembles NGC\,4691 in many aspects. Both have
extremely faint and smooth arms and a chaotic system of dust filaments, in all
the interior of the inner ring for NGC\,1022 \citep{Sandageb}. NGC\,1022 has a
very short bar for a SBa galaxy. All interstellar tracers are highly concentrated
in a small circumnuclear region \citep{Garciab, Hameed} and we detect
only this central region of size $< 10\arcsec \approx 1$\,kpc in the mid-infrared.
The morphology of both galaxies is suggestive of a past merger. 

\item {\bf Magellanic barred spirals:}

NGC\,337 (SBd) and 4027 (SBdm).

These are the two latest-type galaxies of the field subsample. They are
strongly asymmetric, with off-centered bars with respect to the outer isophotes.
One arm harbors a very bright complex of star formation not far from its
connection to the bar (this complex is clearly delineated in our mid-infrared images)
and the second arm is embryonic. In some magellanic barred spirals, the
bar center does not correspond to the kinematic center \citep[but this is
not the case for NGC\,4027, according to][]{Pence}. This morphology is thought to
be driven by tidal interaction \citep{Odewahn}, or can result from the
cooperation between spiral waves of modes $m=1$ and $m=2$ \citep{Tagger}, the
two interpretations not being exclusive.

\end{list}

\subsection{Virgo subsample}

This subsample includes 9 or 10 spirals with normal H{\scriptsize I} content
and 14 or 13 H{\scriptsize I}-deficient spirals (NGC\,4293 is at the limit between
the two categories). H{\scriptsize I}-deficient galaxies do not necessarily belong
to the cluster core, but can be found at large projected distances from M\,87.
For instance, NGC\,4394, 4450, 4498 and 4580, with high H{\scriptsize I}
deficiencies, are located in the cluster periphery, between 1.3 and 2\,Mpc
in projected distance from M\,87.

Seven Virgo members of the sample have low blue or H-band luminosities
\citep[H data from][ used because they give an indication on the total
stellar mass]{Bosellib}, and their sizes are also intrisically small. They
are either normal in their H{\scriptsize I} content (NGC\,4351, 4430 and 4633)
and of rather irregular morphology, except NGC\,4634 seen edge-on, or deficient
with no obvious distorsion (NGC\,4413, 4491, 4498 and 4506).

Interestingly, the galaxies presented in this paper which have a central
$F_{15}/F_7$ ratio above 2 are all Virgo spirals (except NGC\,1022), with large
H{\scriptsize I} deficiencies. We computed the H{\scriptsize I} deficiency
according to \citet{Guiderdoni} for the galaxies in our sample (the values are
given in Paper~I). This quantity is defined as the logarithmic difference in
H{\scriptsize I} mass surface density between a reference field galaxy sample and the
galaxy considered, normalized by the dispersion in the field sample. We have
adopted as the threshold for H{\scriptsize I} dearth $Def > 1.2$ (Paper~I). For all
the galaxies with a central color $F_{15}/F_7 > 2$ except NGC\,1022 (namely
NG\,4569, 4293, 4388 and 4491), $1.2 < Def < 2$.
This suggests that the interaction with the intracluster gas also has
consequences for the internal dynamics and reinforces the effects of the bar.
This is not a systematic effect, since the remaining 10 H{\scriptsize I}-deficient
galaxies have low to moderate central colors.

\section{Optical and mid-infrared disk sizes}
\label{sec:sizes}

\begin{figure}[!b]
\vspace*{6cm}
\caption{Histograms of mid-infrared to optical size ratios as a function of
         H{\scriptsize I}-deficiency and morphological type. $Def > 1.2$
         corresponds to severely H{\scriptsize I}-stripped galaxies and
         $Def < -0.3$ to gas-rich galaxies.}
\label{fig:taille_def}
\end{figure}

We have determined the size of the infrared emitting disk as explained at the
beginning of Sect.~\ref{sec:morpho}, and also quantified the H{\scriptsize I}
deficiency, which is closely related to the anemia phenomenon (star formation being
inhibited in parts of the outer disk where the gas density has dropped due to
stripping). We find that the ratio of the infrared diameter, $D_{5~\mu Jy}^7$, to
the optical diameter $D_{25}$, defined at the blue isophote 25 mag.arcsec$^{-2}$
\citep{Vaucouleurs}, shows marked variations as a function of H{\scriptsize I}
deficiency (Fig.~\ref{fig:taille_def}, top). Whereas $D_{5~\mu Jy}^{7} / D_{25}$
is of the order of 0.7--0.9 for gas-rich spirals (none of the Virgo galaxies belongs
to this category) and 0.96 for NGC\,4634 which is seen edge-on, it ranges between
0.5 and 0.8 for spirals of intermediate H{\scriptsize I} content, and between
0.35 and 0.65 for deficient spirals, except 0.77 for NGC\,4647 whose outskirts
may be contaminated by the emission of the neighbor elliptical NGC\,4649. The
value 0.35 is that of NGC\,4569, known for its spectacular anemia, and NGC\,4438,
which has undergone a violent collision \citep{Combes}. The only exception to
this trend is NGC\,289, which is very rich in H{\scriptsize I} but has a ratio
$D_{5~\mu Jy}^{7} / D_{25}$ of about 0.4. We again insist that its outer disk is
of very low surface brightness, and that the condition for star formation in such
disks is likely to be very different from those in ordinary bright disk galaxies.

There is also a trend for $D_{5~\mu Jy}^{7} / D_{25}$ to be lower in early-type
than in late-type spirals, as shown in Fig.~\ref{fig:taille_def} (bottom).
However, this relationship is not independent of the first one, since
H{\scriptsize I}-deficient galaxies are mostly classified as early, due to the
dominance in practice of the criterion based on the aspect of the arms in the
Hubble-system classification (see Paper~I). Early-type spirals are essentially
characterized by smooth and faint outer disks, whereas in latest types, the arms
are bright and patchy in their whole length, structured by star formation sites.
Hence, a natural interpretation of both histograms shown here is that the ratio of
mid-infrared to optical sizes primarily depends on the star formation activity
in outer disks, which is strongly coupled to the gas density.

\section{Summary}

We have presented complete maps of dust emission at 7\,$\mu$m of 43 galaxies
spanning the spiral sequence from types S0/a to Sdm, at an angular resolution
less than $10\arcsec$ and a sensitivity of the order of $5\,\mu$Jy.\,arcsec$^{-2}$.
We have also detailed the data reduction, provided the total fluxes at 7 and
15\,$\mu$m and outlined the generic morphological properties in this sample.
Since there exists an abundant literature on many of these galaxies studied
individually, the present atlas has concentrated on common features.
The detailed description of the sample and interpretation of the data are
given together in Paper~I.

The morphology is very similar in the mid-infrared and in the optical,
except in peculiar galaxies where the dust emission is highly concentrated:
the 7 and 15\,$\mu$m emission follows tightly spiral arms, lenses and
individual giant complexes. It appears however less extended than the stellar
emission and tends to be reduced in smooth and unstructured spiral arms
often seen in the outer disks of early-type galaxies and especially of
H{\scriptsize I}-deficient galaxies. This can be related to low densities of
both the gas and the young stellar population.
Another noticeable difference concerns the aspect of the bar, where the responses
of the interstellar tracers and of the stellar component to the gravitational
potential are clearly distinct.
Finally, central regions stand out in our maps as entities with particular
properties, and are studied in detail in Paper~I.

The major characteristic of mid-infrared colors, {\it i.e.} 15\,$\mu$m to
7\,$\mu$m flux ratios, is that they are remarkably uniform throughout disks,
except in some circumnuclear regions and a few bright complexes in spiral arms
(the best example of the latter can be found in the outer interacting arms of
M\,101).

The maps will be made available to the community, in fits format,
in the ISO public archive (http://www.iso.vilspa.esa.es/). They can be used for
studies of the interplay between dust and other components of galaxies,
and constitute the basis for two following papers.

\begin{table*}[!t]
\caption[]{Raster observations.
           Listed are the number of the figure showing the maps, the project name
           (B: {\it Cambarre}; S: {\it Camspir}; F: {\it Sf\_glx}; V: {\it Virgo}),
           the pixel size, the number of
           pointings projected onto the raster map, the field of view, the number
           of exposures per pointing, the integration time per exposure, total
           on-target times at 15 and 7\,$\mu$m, and total fluxes.
           When two figures are given, the first one corresponds to 15\,$\mu$m
           and the second one to 7\,$\mu$m.
           For NGC\,6946, observed with $6\arcsec$ pixels, each pointing
           is displaced from the previous by a fractional number of pixels,
           so that the pixel size of the projected map is $3\arcsec$.}
\label{tab:tab_rasters}
\begin{minipage}[t]{15cm}
\begin{flushleft}
\begin{tabular}{|llccrcccrrr@{$\,$}lr@{$\,$}l|}
\hline
\noalign{\smallskip}
name & fig. & proj. & pix. & N$_{\rm p}$ & map size &
   n$_{\rm exp}$\footnote{This is the minimum number of exposures.
   At the beginning of the observation for each filter, it is in general much larger,
   in order to get closer to stabilization. \\
   \hspace*{1ex} \dag\ The map size given is that of the combination of the main map
   with the eastern arm map.} &
   t$_{\rm int}$ & T$_{15}$ & T$_7$ &
   \multicolumn{2}{c}{$F_{15}$} & \multicolumn{2}{c|}{$F_7$} \\
~ & ~ & ~ & ($\arcsec$) & ~ & ($\arcsec$) & ~ & (s) & \multicolumn{2}{c}{(s)} &
   \multicolumn{4}{c|}{(mJy)} \\
\noalign{\smallskip}
\hline
\noalign{\smallskip}
N289                   & 6c   & B & 3	& 16  & 240		  & 26     & 5.0      & 2086. & 2086. &   327.8 $\pm$ &   25.6 &   342.9$\pm$ &  14.7 \\
N337                   & 7b   & B & 3	& 16  & 240		  & 26     & 5.0      & 2131. & 2137. &   297.9 $\pm$ &   24.0 &   336.1$\pm$ &  17.9 \\
N613                   & 5c   & B & 3	& 25  & 288		  & 25     & 5.0      & 3170. & 3170. &  1566.5 $\pm$ &  104.0 &  1473.3$\pm$ &  71.4 \\
N1022                  & 7a   & B & 3	& 16  & 240		  & 26     & 5.0      & 2071. & 2076. &   802.3 $\pm$ &   86.4 &   444.4$\pm$ &  45.3 \\
N1097                  & 5d   & B & 3	& 25  & 288		  & 25     & 5.0      & 3119. & 3114. &  2269.2 $\pm$ &  167.4 &  2128.6$\pm$ & 125.4 \\
N1365                  & 5d   & S & 6	& 16  & 480		  & 61     & 2.1      & 1978. & 1984. &  4436.7 $\pm$ &  764.5 &  3691.9$\pm$ & 616.6 \\
N1433                  & 5b   & B & 3	& 25  & 288		  & 25     & 5.0      & 3165. & 3175. &   355.3 $\pm$ &   41.0 &   381.3$\pm$ &  33.8 \\
N1530                  & 5b   & B & 3	& 16  & 240		  & 26     & 5.0      & 2137. & 2126. &   606.1 $\pm$ &   39.2 &   573.9$\pm$ &  39.1 \\
N1672                  & 5c   & B & 3	& 25  & 288		  & 25     & 5.0      & 3170. & 3155. &  2020.5 $\pm$ &  123.0 &  1985.0$\pm$ & 129.2 \\
N4027                  & 7b   & B & 6	& 9   & 384		  & 41, 42 & 2.1      &  787. &  787. &   676.7 $\pm$ &   95.5 &   775.8$\pm$ &  68.2 \\
~                      & ~    & ~ & 3	& 4   & 132		  & 42, 41 & 2.1      &  350. &  359. & \multicolumn{4}{c|}{~} \\
N4535                  & 6b   & B & 6	& 16  & 480		  & 27, 20 & 5.0      & 2066. & 2056. &  1127.9 $\pm$ &  181.4 &  1136.6$\pm$ &  68.9 \\
~                      & ~    & ~ & 3	& 4   & 132		  & 26     & 5.0      &  539. &  534. & \multicolumn{4}{c|}{~} \\
N4691                  & 7a   & B & 6	& 4   & 288		  & 41, 42 & 2.1      &  359. &  357. &   795.9 $\pm$ &  185.6 &   613.5$\pm$ &  83.1 \\
~                      & ~    & ~ & 3	& 4   & 132		  & 42, 41 & 2.1      &  352. &  354. & \multicolumn{4}{c|}{~} \\
N4736 (M94)            & 6d   & S & 6	& 9   & 288		  & 20     & 5.0      &  917. &  922. &  4204.5 $\pm$ &  240.6 &  3913.9$\pm$ & 225.8 \\
N5194 (M51)            & 6d   & S & 3	& 100 & 663		  & 27     & 2.1      & 5384. & 5390. &  8003.2 $\pm$ &  493.5 &  8598.7$\pm$ & 552.1 \\
N5236 (M83)            & 6a   & S & 6	& 49  & 768		  & 21     & 5.0      & 4918. & 5155. & 20098.4 $\pm$ &  803.7 & 18474.9$\pm$ & 899.7 \\
N5383                  & 5a   & B & 6	& 9   & 384		  & 54     & 2.1      & 1024. & 1022. &   332.6 $\pm$ &   61.9 &   350.2$\pm$ &  62.1 \\
~                      & ~    & ~ & 3	& 4   & 156		  & 54, 52 & 2.1, 5.0 &  459. & 1083. & \multicolumn{4}{c|}{~} \\
\multicolumn{14}{|l|}{N5457 (M101)} \\
\hspace{0.3cm} main \dag & 6c & S & 6	& 49  & $|~1302$~         & 20, 21 & 5.0      & 4949. & 4924. &  5424.3 $\pm$ &  322.0 &  6034.0$\pm$ & 116.7 \\
\hspace{0.3cm} E arm \dag & ~ & ~ & 6	& 25  & $|\times 972$     & 25     & 5.0      & 3155. & 3140. & \multicolumn{4}{c|}{~} \\
N6744                  & 6b   & S & 6	& 16  & 480		  & 25     & 5.0      & 2031. & 2041. &  1497.4 $\pm$ &  125.7 &  2419.4$\pm$ &  52.3 \\
N6946                  & 6a   & F & 6-3 & 64  & 759		  & 13, 9  & 2.1, 5.0 & 1585. & 2872. & 10651.6 $\pm$ & 1767.2 & 11648.8$\pm$ & 678.6 \\
N7552                  & 5a   & B & 6	& 16  & 480		  & 26     & 5.0      & 2091. & 2112. & \multicolumn{4}{c|}{~} \\
~                      & ~    & ~ & 3	& 9   & 168		  & 26     & 5.0      & 1184. & 1174. &  2767.6 $\pm$ &  193.7 &  1826.2$\pm$ & 168.5 \\
\multicolumn{14}{|l|}{\it Virgo cluster sample~:} \\
\multicolumn{14}{|l|}{~} \\
N4178                  & 8a   & V & 6	& 25  & 576		  & 17, 18 & 2.1      &  819. &  860. &   181.5 $\pm$ &   48.0 &   228.5$\pm$ &  24.6 \\
N4192                  & 8a   & V & 6	& 49  & 768		  & 17, 18 & 2.1      & 1528. & 1625. &   630.0 $\pm$ &   99.6 &   900.8$\pm$ &  68.3 \\
N4293                  & 9a   & V & 6	& 25  & 576		  & 17, 18 & 2.1      &  804. &  844. &   188.6 $\pm$ &   42.8 &   159.5$\pm$ &  25.3 \\
N4351                  & 8b   & V & 6	& 9   & 384		  & 17, 15 & 2.1      &  291. &  283. &    45.6 $\pm$ &   26.3 &    52.6$\pm$ &   8.7 \\
N4388                  & 9a   & V & 6	& 25  & 576		  & 17, 18 & 2.1      &  812. &  863. &  1008.2 $\pm$ &  244.0 &   499.4$\pm$ &  77.8 \\
N4394                  & 9b   & V & 6	& 16  & 480		  & 16     & 2.1      &  480. &  478. &   139.0 $\pm$ &   41.0 &   161.2$\pm$ &  19.1 \\
N4413                  & 9b   & V & 6	& 9   & 384		  & 17, 16 & 2.1      &  302. &  281. &    93.0 $\pm$ &   31.4 &    89.3$\pm$ &  11.0 \\
N4430                  & 8b   & V & 6	& 9   & 384		  & 17, 16 & 2.1      &  294. &  279. &    98.0 $\pm$ &   23.5 &   132.5$\pm$ &  13.9 \\
N4438                  & 9c   & V & 6	& 49  & 768		  & 17, 18 & 2.1      & 1528. & 1629. &   209.1 $\pm$ &   34.7 &   231.9$\pm$ &  26.9 \\
N4450                  & 9c   & V & 6	& 25  & 576		  & 17     & 2.1      &  791. &  844. &   169.7 $\pm$ &   42.5 &   185.1$\pm$ &  14.6 \\
N4491                  & 9d   & V & 6	& 9   & 384		  & 17, 15 & 2.1      &  310. &  287. &    81.1 $\pm$ &   25.2 &    30.5$\pm$ &   7.6 \\
N4498                  & 9d   & V & 6	& 16  & 480		  & 16     & 2.1      &  493. &  478. &    94.6 $\pm$ &   19.2 &   112.9$\pm$ &  11.8 \\
N4506                  & 9e   & V & 6	& 9   & 384		  & 17, 16 & 2.1      &  294. &  277. &    12.7 $\pm$ &    5.2 &    21.1$\pm$ &   9.8 \\
N4567/                 & 8c   & V & 6	& 25  & 576		  & 17, 18 & 2.1      &  779. &  829. &   293.4 $\pm$ &   15.5 &   317.9$\pm$ &  16.4 \\
\hspace{0.3cm} N4568   & \multicolumn{9}{c}{~}                                                        &  1099.0 $\pm$ &  127.6 &  1074.7$\pm$ &  64.8 \\
N4569                  & 9e   & V & 6	& 64  & 864		  & 16, 18 & 2.1      & 1864. & 2108. &   939.3 $\pm$ &  125.1 &   843.5$\pm$ &  54.1 \\
N4579                  & 9f   & V & 6	& 36  & 672		  & 17, 19 & 2.1      & 1161. & 1306. &   619.2 $\pm$ &   85.1 &   672.5$\pm$ &  37.5 \\
N4580                  & 9f   & V & 6	& 9   & 384		  & 17, 16 & 2.1      &  298. &  279. &   103.9 $\pm$ &   24.2 &   102.6$\pm$ &   7.7 \\
N4633/                 & 8c/  & V & 6   & 36  & 672		  & 17, 19 & 2.1      & 1123. & 1266. &    30.0 $\pm$ &    9.5 &    30.3$\pm$ &   9.1 \\
\hspace{0.3cm} N4634   & 8d   & \multicolumn{8}{c}{~}                                                 &   258.2 $\pm$ &   40.7 &   278.3$\pm$ &  35.0 \\
N4647                  & 9g   & V & 6	& 36  & 672		  & 17, 18 & 2.1      & 1127. & 1310. &   472.3 $\pm$ &   32.0 &   474.3$\pm$ &  17.2 \\
N4654                  & 8d   & V & 6	& 25  & 576		  & 17, 18 & 2.1      &  814. &  852. &  1018.6 $\pm$ &   78.4 &  1049.4$\pm$ &  42.9 \\
N4689                  & 9g   & V & 6	& 25  & 576		  & 17, 18 & 2.1      &  787. &  863. &   329.7 $\pm$ &   37.4 &   340.9$\pm$ &  16.3 \\
\noalign{\smallskip}
\hline
\end{tabular}
\end{flushleft}
\end{minipage}
\end{table*}

\begin{table*}[!t]
\caption[]{Parameters of the spectral observations.
           They include the pixel size, the ratio of the field of view
	   to that of raster maps, the number of exposures per
	   wavelength, the integration time per exposure and the
	   total useful time.}
\label{tab:tab_cvfs}
\begin{minipage}[t]{15cm}
\begin{flushleft}
\begin{tabular}{|lccccc|}
\hline
\noalign{\smallskip}
name & pix. size ($\arcsec$) & frac. size &
n$_{\rm exp}$ & t$_{\rm int}$ (s) & T$_{\rm tot}$ (s) \\
\noalign{\smallskip}
\hline
\noalign{\smallskip}
N613\footnote{At $3\arcsec$, no calibration flat-field in narrow filters was
available, so we did not apply the division by $ff_{\rm filter}$ (see last
paragraph of Section~\ref{sec:cvf}).}
      & 3 & 0.33 & 12    & 2.1 & 3960. \\
N1097\footnote{At the change of variable filter at 9.2\,$\mu$m, a downward jump
was visible in the whole field; to cancel it, we applied a unique offset to
the second part of the spectrum.}
      & 6 & 0.66 & 9--10 & 2.1 & 3276. \\
N1365 & 6 & 0.40 & 15    & 2.1 & 4918. \\
N5194 & 6 & 0.29 & 15    & 2.1 & 4918. \\
N5236 & 6 & 0.25 & 15    & 2.1 & 4918. \\
\noalign{\smallskip}
\hline
\end{tabular}
\end{flushleft}
\end{minipage}
\end{table*}

\clearpage

\noindent
Fig.~5:
Strongly barred galaxies by order of decreasing bar strength.
         {\bf a:} NGC\,5383 and 7552, with $D_{\rm bar}/D_{25} \approx 0.65$ and
         0.67. In all figures are displayed, from top to bottom: an optical image
         from the DSS with, in a corner, a segment showing the orientation of the
         line of nodes and the inclination value; the 7\,$\mu$m map; $F_{15}/F_7$
         colors of selected regions (`C': circumnuclear region; `D': averaged disk;
         'Cr': central resolution element). Locations in the bar (`B') or the arms
         (`A') are indicated on the 7\,$\mu$m map, which is scaled identically to
         the optical map except when a dashed segment outlines the field of view
         shown below.
\vspace*{0.5cm}

\noindent
Fig.~5:
{\bf b:} NGC\,1433 and 1530, with $D_{\rm bar}/D_{25} \approx 0.54$ and
         0.48.
\vspace*{0.5cm}

\noindent
Fig.~5:
{\bf c:} NGC\,613 and 1672, with $D_{\rm bar}/D_{25} \approx 0.45$ and
         0.42.
\vspace*{0.5cm}

\noindent
Fig.~5:
{\bf d:} NGC\,1097 and 1365, with $D_{\rm bar}/D_{25} \approx 0.35$ and
         0.31. The labels `S' and `N' correspond to the brightest spot in the ring
         and the nucleus, whose colors were estimated from the images treated with
         the analog of CLEAN (Sect.~\ref{sec:sepcnr}), with no error bars.
\vspace*{0.5cm}

\noindent
Fig.~6:
Weakly barred and unbarred galaxies.
         {\bf a:} NGC\,5236 and 6946, moderately barred with
         $D_{\rm bar}/D_{25} \approx 0.24$ and 0.19.
\vspace*{0.5cm}

\noindent
Fig.~6:
{\bf b:} NGC\,4535 and 6744, weakly barred with
         $D_{\rm bar}/D_{25} \approx 0.18$ and 0.17.
\vspace*{0.5cm}

\noindent
Fig.~6:
{\bf c:} NGC\,289 and 5457, with $D_{\rm bar}/D_{25} \approx 0.15$ and
         0.06.
\vspace*{0.5cm}

\noindent
Fig.~6:
{\bf d:} NGC\,4736 and 5194, unbarred galaxies (NGC\,4736 possesses in fact
         an oval distorsion). The labels `N' and `Cp' refer to the nucleus of
         NGC\,5194 and the nucleus of its companion, NGC\,5195.
         Note that in NGC\,5194, the thin NE and SW outermost
         parts of both arms seen in the mid-infrared coincide with dust lanes
         seen in absorption.
\vspace*{0.5cm}

\noindent
Fig.~7:
Peculiar galaxies.
         {\bf a:} NGC\,1022 and 4691, amorphous barred spirals (possibly merger
         results). They have $D_{\rm bar}/D_{25} \approx 0.32$ and 0.43. The labels
         `S1' and `S2' refer to the two resolved spots in the center of NGC\,4691,
         whose colors were estimated from the images treated with the analog of
         CLEAN, with no error bars.
\vspace*{0.5cm}

\noindent
Fig.~7:
{\bf b:} NGC\,337 and 4027, magellanic barred spirals. The bar length was
         not estimated because it is difficult to delineate the bar precisely.
\vspace*{0.5cm}

\noindent
Fig.~8:
Virgo galaxies which are not H{\scriptsize I}-deficient.
         {\bf a:} NGC\,4178 and 4192.
\vspace*{0.5cm}

Fig.~8:
{\bf b:} NGC\,4351 and 4430.
\vspace*{0.5cm}

\noindent
Fig.~8:
{\bf c:} NGC\,4567/8 and 4633.
\vspace*{0.5cm}

\noindent
Fig.~8:
{\bf d:} NGC\,4634 and 4654.
\vspace*{0.5cm}

\noindent
Fig.~9:
H{\scriptsize I}-deficient Virgo galaxies.
         {\bf a:} NGC\,4293 and 4388.
\vspace*{0.5cm}

\noindent
Fig.~9:
{\bf b:} NGC\,4394 and 4413.
\vspace*{0.5cm}

\noindent
Fig.~9:
{\bf c:} NGC\,4438 and 4450. The label `Cp' refers to the nucleus of
         NGC\,4435, which is likely at the origin of the disruption of NGC\,4438
         \citep{Combes}.
\vspace*{0.5cm}

\noindent
Fig.~9:
{\bf d:} NGC\,4491 and 4498.
\vspace*{0.5cm}

\noindent
Fig.~9:
{\bf e:} NGC\,4506 and 4569.
\vspace*{0.5cm}

\noindent
Fig.~9:
{\bf f:} NGC\,4579 and 4580.
\vspace*{0.5cm}

\noindent
Fig.~9:
{\bf g:} NGC\,4647 and 4689.

\clearpage

\begin{acknowledgements}
~
\vspace*{1ex} \\
The DSS images shown here were retrieved from the ESO/ST-ECF Archive and were
produced at the STScI. They are based on photographic data obtained using the
Oschin Schmidt Telescope operated by Caltech and the Palomar Observatory, and the
UK Schmidt Telescope operated by the Edinburgh Royal Observatory.
\vspace*{1ex} \\
The ISOCAM data presented in this paper were analyzed using and adapting the CIA
package, a joint development by the ESA Astrophysics Division and the ISOCAM
Consortium (led by the PI C. Cesarsky, Direction des Sciences de la Mati\`ere,
C.E.A., France).
\end{acknowledgements}

\bibliographystyle{apj}

\end{document}